\documentclass[runningheads]{llncs}
\usepackage[T1]{fontenc}
\usepackage{graphicx}
\usepackage[colorlinks=true]{hyperref}
\usepackage{color}
\hypersetup{
    linkcolor=blue,  % Color of internal links (\ref, \eqref, table of contents)
    citecolor=blue,  % Color of citations (\cite)
    urlcolor=blue    % Color of web links (\url, \href)
}

\usepackage[caption=false]{subfig}
\usepackage{multirow}
\usepackage{pifont}
\usepackage{amsfonts}
\usepackage{listings}
\usepackage{booktabs}
\usepackage{float}
\usepackage[nocompress]{cite}

\begin{document}
\title{Large-Scale User Behavior Analysis in \\ Multimodal AI-Assisted Manual Task Execution}

\titlerunning{User Behavior Analysis in Multimodal AI-Assisted Manual Task Execution}

\author{Rafael Ferreira%\orcidID{0000-0002-2086-1356} 
\and
Diogo Tavares%\orcidID{0000-0002-0147-369X} 
\and
Diogo Glória-Silva%\orcidID{0000-0002-4420-7455} 
\and \\
David Semedo%\orcidID{0000-0002-2403-0058} 
\and
João Magalhães%\orcidID{0000-0001-6290-5719}
}
% %
\authorrunning{Ferreira, R. et al.}
\institute{NOVA University of Lisbon, Portugal, 
NOVA LINCS, Portugal \\
\email{\{rah.ferreira, dc.tavares, dmgc.silva\}@campus.fct.unl.pt} \\
\email{\{df.semedo, jmag\}@fct.unl.pt}
}

\maketitle              % typeset the header of the contribution
\begin{abstract}

Conversational Task Assistants (CTAs) are multimodal dialogue systems that support users in complex real-world tasks such as cooking and DIY through voice, text, image, and video interactions.
Prior user studies have focused on controlled settings, leaving limited understanding of real-world CTA usage at scale.
In this work, we present a large-scale study of CTA usage based on thousands of users in-the-wild.
Our large-scale real-world data analysis unveils new understandings of (i) user-CTA interaction flows, (ii) user intents, (iii) user conversational traits, and (iv) behavioral factors associated with user satisfaction.
Our findings reveal key opportunities for future research in CTAs, particularly in user interaction design and task engagement, concluding with concrete design guidelines.


\end{abstract}
\section{Introduction}
\label{sec_intro}

Supporting users through real-world tasks such as cooking and DIY remains a challenge for conversational AI.
Unlike traditional task-oriented dialogue systems that focus on transactional interactions such as booking or recommendation~\cite{multiwoz_dataset,tod_systems_survey,mmwoz_dataset}, Conversational Task Assistants (CTAs) are multimodal systems designed to support open-ended, mixed-initiative interactions across multi-step tasks~\cite{taskbot_overview_year_2, wizard_of_tasks, task2dial, grillbot_sigkdd}.

Despite growing interest in CTAs, most existing work evaluates these systems in controlled environments or crowdsourced settings~\cite{grillbot_sigkdd, cooking_taskbot_chi, mango_mango}. While useful for controlled experimentation, these settings often fail to capture the variability and dynamics of real users in deployed systems~\cite{tavares_fake_users, wang2025human}. As a result, how users interact with tasks in real-world CTA deployments remains understudied.

In this work, we address this gap through a large-scale user study of TWIZ-v2~\cite{twiz_v2}, a CTA deployed in a production environment that provides structured task guidance via multimodal interaction (voice, text, images, and video).
In particular, we analyze user interactions collected in the context of the Alexa TaskBot Challenge~\cite{taskbot_overview_year_2}, where the system was available to the entire U.S. Alexa user base in a real-world setting. 
This setting enables observation of natural user behavior at scale, including task discovery, step-by-step interaction patterns, and task engagement. 
Additionally, a subset of these interactions includes 1–5 user ratings, enabling analysis of satisfaction signals.

Therefore our goal is to understand how users interact with CTAs in the real-world, as shown in Figure~\ref{fig_conversation_example}. 
Specifically, our contributions are as follows: \textbf{(1)} we present a large-scale analysis of user–CTA interaction flows, examining how users initiate and progress through various stages; \textbf{(2)} we analyze user intents across conversations; \textbf{(3)} we study user conversational traits and compare real-world interactions with existing crowdsourced CTA datasets; and \textbf{(4)} we identify behavioral signals associated with user satisfaction in CTA systems.

Our findings reveal opportunities to improve CTAs, particularly in handling uncooperative user behavior, improving task discovery, and supporting sustained engagement during task execution.
Overall, this analysis provides concrete insights for designing more effective multimodal conversational task assistants.

\section{Related Work}
\label{sec_related_work}

\subsubsection{Conversational Task Assistants.}
\label{sub_rw_ctas}
 
To support research in the CTA domain, crowdsourced datasets have been introduced. In particular, Wizard of Tasks~\cite{wizard_of_tasks} and Task2Dial~\cite{task2dial} provide task-centric dialogues addressing challenges such as intent recognition and commonsense reasoning. However, these datasets do not fully capture the variability of real-world usage, highlighting a gap between crowdsourced and deployed systems~\cite{tavares_fake_users}.

Among efforts to collect real-world CTA data, the Alexa TaskBot Challenge~\cite{taskbot_overview_year_2} focuses on CTA for cooking and DIY tasks. The challenge emphasizes robustness in real user conditions, requiring systems to handle multimodal data. Notable systems include GRILLBot~\cite{grillbot_sigkdd}, which demonstrated the effectiveness of hybrid architectures combining LLMs with specialized components for task guidance. Other systems, such as our CTA TWIZ~\cite{twiz_v2} and ISABEL~\cite{isabel_taskbot}, further explore multimodal task assistance by integrating text, images, and video to provide context-aware support.
However, no fine-grain analysis of these interactions has been performed.

\subsubsection{User Behavior with CTAs.}
\label{sub_rw_user_behavior}

Prior work highlights the importance of natural, proactive conversational strategies~\cite{mango_mango}, as well as the role of multimodal assistants in improving user autonomy and task understanding~\cite{cooking_bot_interaction}. User studies have also identified limitations, especially in delivering context-aware and timely responses, motivating the need for more adaptive dialogue strategies~\cite{cooking_taskbot_chi, cooking_in_conversation}.

Visual aids also have a critical role in task comprehension~\cite{media_in_tasks}, while also reducing cognitive load during interactions~\cite{cooking_taskbot_chi}. This is particularly important for voice-based systems, where users rely heavily on working memory, motivating the adaptation of content for spoken interfaces~\cite{adapting_recipes_to_ctas}. These challenges are further amplified in noisy, real-world environments where ASR errors are common, underscoring the need for resilient, error-tolerant design~\cite{surpassing_asr_errors}.

Building on these findings, our work goes beyond controlled studies to analyze large-scale, real-world user interactions, providing insights into CTA usage and behavior under everyday deployment conditions.

\begin{figure*}[tb]
\centering
\includegraphics[width=1.00\linewidth]{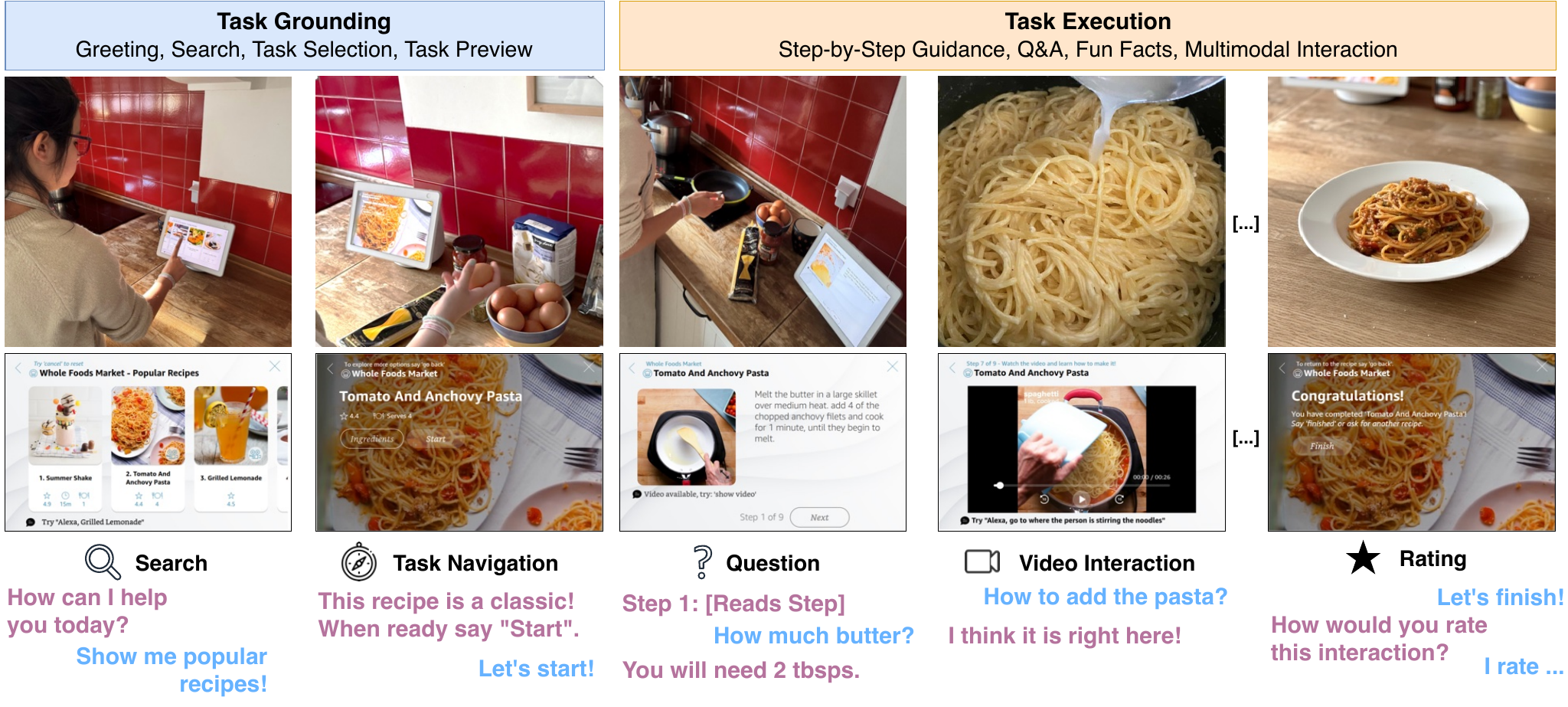}
\caption{Illustrative conversation between a user and our CTA system showcasing the multi-stage nature of the interaction and its various conversation sub-flows.}
\label{fig_conversation_example}
\end{figure*}

\section{Designing CTAs for Effective Task Guidance}
\label{sec_system_design}

In this section, we briefly introduce TWIZ-v2\footnote{See~\cite{twiz_v2} for comprehensive implementation details of all system components.}, the multimodal CTA that enabled the data collection for this user study.
The system is designed around three principles: (1) modular architecture, (2) multimodal integration, and (3) context-aware responses.

\subsection{Modular System Architecture}
\label{sub_system_architecture}
\begin{figure}[tb]
    \begin{minipage}[b]{0.48\textwidth}
        \centering
        \includegraphics[width=\textwidth]{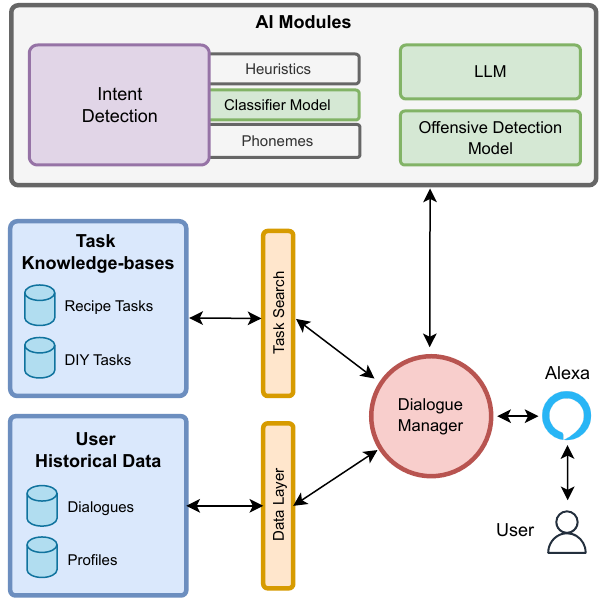}
        \caption{The system's modular architecture.}
        \label{fig_system_architecture}
    \end{minipage}\hfill
    \begin{minipage}[b]{0.45\textwidth}
        \centering
        \includegraphics[width=\textwidth]{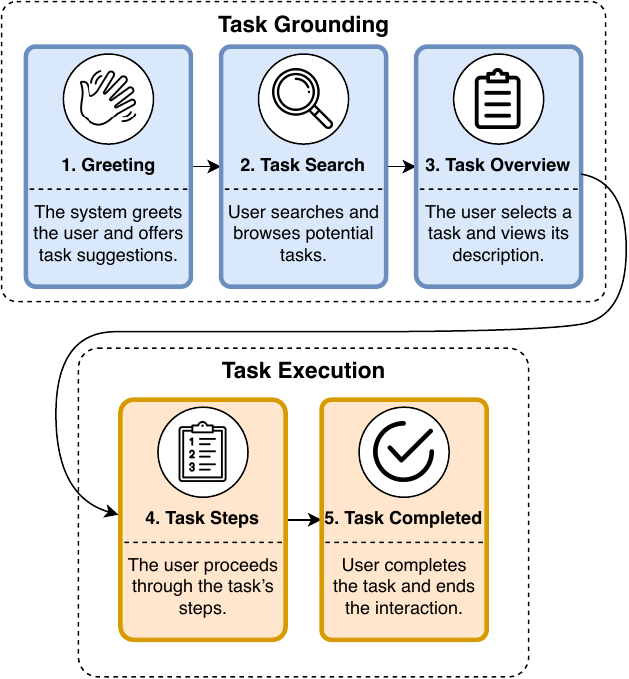}
        \caption{Dialogue manager interaction stages and sub-stages.}
        \label{fig_dialogue_stages}
    \end{minipage}
\end{figure}

To support complex CTA interactions, the system is organized into modular components for managing dialogue flow and task content, as illustrated in Figure~\ref{fig_system_architecture}. The system is built using the Cobot~\cite{cobot} framework, where a serverless function orchestrates dialogue management and access to other modules. Conversation history and task search are handled by a database and a search system, respectively.

\subsubsection
{Dialogue Management.}
\label{sub_sub_dialogue_manager}

The dialogue manager follows an event-driven, stage-based state machine that tracks both dialogue and task progress. Figure~\ref{fig_dialogue_stages} illustrates the interaction stages and their roles which are organized into two macro-stages, each composed of several sub-stages:

\begin{itemize}
\item \textbf{Task Grounding}, which establishes context and supports task selection, consists of the \textit{Greeting}, \textit{Task Search}, and \textit{Task Overview} sub-stages.
\item \textbf{Task Execution}, which guides users through the selected task, consists of the \textit{Task Steps} and \textit{Task Completed} sub-stages.
\end{itemize}

This structure provides clear interaction guidance and helps users remain aware of their current position within the task flow.

To support flexible interactions, the system includes sub-flows for Q\&A and multimodal interaction, allowing users to temporarily diverge from the main flow while preserving dialogue continuity as shown in the example of Figure~\ref{fig_conversation_example}.

\subsubsection{Intent Detection.}
\label{sub_sub_intent_detection}

For each user utterance, the system identifies the user’s intent and uses it to drive system behavior (full intent set in Figure~\ref{fig_intent_distribution}). Here, accuracy is critical as ASR errors are common in voice-based systems~\cite{surpassing_asr_errors}.
To improve robustness, the system combines multiple complementary signals using a voting-style decision mechanism, integrating a BERT-based neural intent classifier, phoneme-level similarity signals~\cite{g2p_intent}, and rule-based heuristics.

\subsubsection{Task Suggestions and Search.}
\label{sub_sub_task_seearch}

To facilitate discoverability~\cite{discovering_commands}, the system proactively provides voice and on-screen task suggestions during the interaction, while also supporting free-form search. Task retrieval is implemented using a Hybrid Search strategy combining text and semantic search. 
Task content is sourced from WikiHow (DIY) and Whole Foods Market (recipes), covering thousands of diverse tasks across several domains.

\subsection{Designing CTAs for Multimodal Dialogues}
\label{sub_ui_navigation}

Given the complexity of tasks, clear and structured guidance is essential for effectively supporting users throughout task execution. To facilitate this, the system incorporates several mechanisms.

\subsubsection{Adaptive Task Chunking.}
\label{sub_sub_breaking_tasks}

To reduce cognitive load, task instructions are broken into shorter steps based on length and formatting cues, omitting non-essential information~\cite{ adapting_recipes_to_ctas}. This is particularly important in voice-only interactions, where users rely on their working memory~\cite{short_term_4}.

\subsubsection{UI and Navigation.}
\label{sub_sub_multimodal_systems}
Users interact via voice in audio-only settings or with combined voice and touch in multimodal environments. Contextual on-screen elements (e.g. images and videos) are synchronized with each dialogue stage, and the UI adopts a design that emphasizes clarity and ease of use (Figure~\ref{fig_conversation_example}).

\subsubsection{Interaction with Multimodal Content.}
\label{sub_sub_multimodal_content}
Users can actively engage with multimodal content. Task-related questions are answered using an LLM-based approach~\cite{vicuna2023} prompted with the current dialogue context. 
Users can also request fun facts~\cite{fun_fact_amazon} or control video playback. 
Video control retrieves relevant video frames by computing the similarity of user queries against cross-modal textual and visual embeddings of keyframes~\cite{clip}.

\subsection{Contextual Dialogue and Response Design}
\label{sub_response_design}

Response design balances clarity and engagement by controlling utterance length and variation. Responses are kept concise to reduce cognitive load~\cite{cooking_taskbot_chi}, commands are placed at the end to aid recall~\cite{discovering_commands}, and phrasing diversity is introduced to reduce repetition. 
Responses are generated using a combination of templates and an LLM, with prompts designed to produce concise and contextual outputs.

To ensure safe interactions, the system detects sensitive topics using heuristics and a RoBERTa-based~\cite{roberta} offensive language classifier. When such topics arise, predefined safety responses are issued~\cite{taskbot_overview_year_2}. For unrecognized requests, the system performs conversation recovery by providing stage-aware contextual explanations and suggesting alternative actions~\cite{discovering_commands}. Generative responses further improve recovery by producing natural, context-aware replies.

\section{Large-scale User Interaction Analysis}
\label{sec_dataset}

In this section, we analyze dialogues with TWIZ-v2 from multiple perspectives, including: (1) user-CTA interaction flows, (2) user intent distribution, (3) user conversational traits, and (4) factors driving user satisfaction.

\subsubsection{Data Collection and Filtering}
\label{sub_data_collection_and filtering}

Our analysis considers interactions with TWIZ-v2~\cite{twiz_v2} over a 12-month period. The system supports manual, multi-step tasks in the cooking and DIY domains via multimodal interactions (voice and visual) and is available to all U.S. Alexa users.
This setup enables the collection of in-the-wild interactions reflecting genuine user behavior. At the end of the interaction, users are prompted to optionally rate the conversation on a 1–5 Likert scale, providing feedback for system evaluation.

To focus on meaningful interactions, we only consider dialogues with at least three turns and removed sessions where over 10\% of turns were unrelated to CTA functionality (e.g., music, or smart home control). The resulting data includes 30k conversations, of which over 5k (16.7\%) received user ratings, a positive conversion rate enabled by explicit feedback request at the end of the interaction. 
To the best of our knowledge, this represents the largest real-world CTA dataset.

\subsubsection{Device and Domain.}
\label{sub_sub_device_domain}
Conversations are from 92\% unique users, with 91\% on screen-equipped devices and 9\% on voice-only devices, showing the predominance of multimodal interactions and capturing a broad spectrum of user behaviors.

Regarding domain selection, among users who select a task (56\% of the total), 54\% engage with recipes and 46\% with DIY tasks, highlighting diverse user interests and the importance of strong performance in both domains.

\begin{figure}[tb]
\centering
\includegraphics[width=0.84\linewidth]{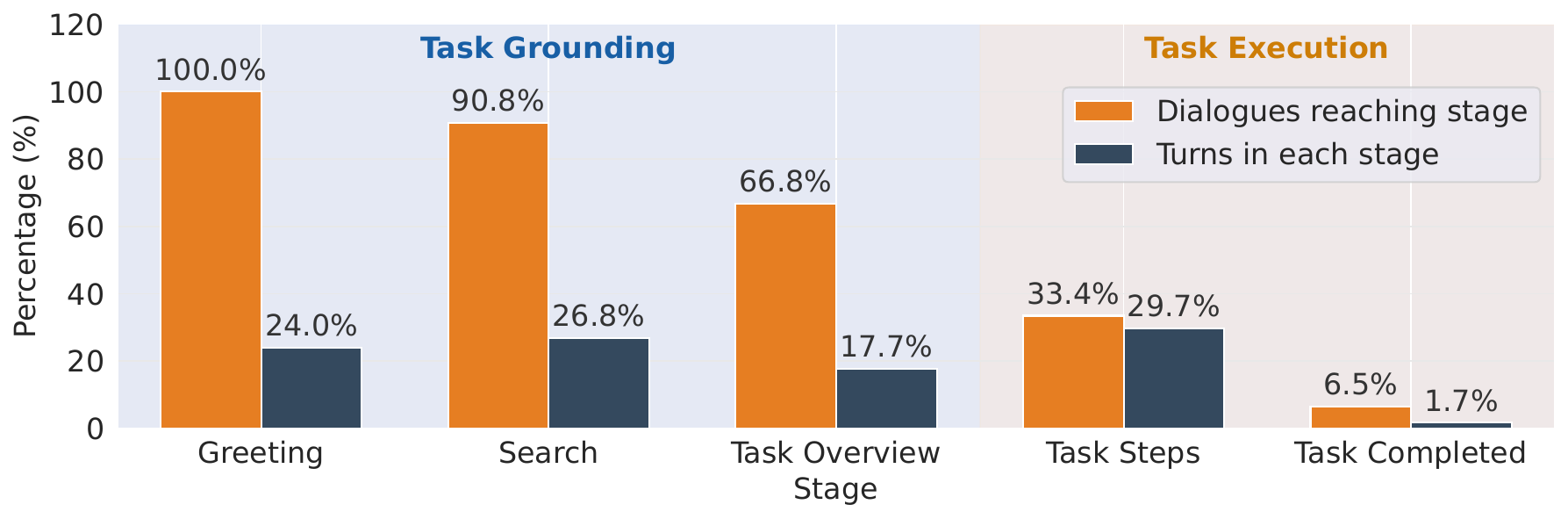}
\caption{Percentage of dialogues reaching a stage and Percentage of turns in each stage.}
\label{fig_stages_distribution}
\end{figure}

\subsection{User Progression Through Dialogue Stages}
\label{sub_sub_stage_based_interaction}

As the system adopts a stage-based interaction pattern that enables tracking user progress, Figure~\ref{fig_stages_distribution} shows both the percentage of dialogues reaching each stage and the distribution of turns in each stage.

As expected, the number of \textbf{dialogue sessions reaching each stage} decreases from the initial to the final stages.
This is particularly evident in the transition from \textit{Search/Task Overview} to \textit{Task Steps}.
This likely indicates that users are primarily exploring the system or evaluating available tasks rather than actively pursuing task completion.
A smaller proportion of users progress through all stages, suggesting opportunities to further support sustained engagement throughout the interaction.

Analyzing the \textbf{turns in each stage}, the \textit{Task Steps} stage accounts for 29.7\% of all turns, making it the most interaction-intensive stage.
The \textit{Search} stage follows as users browse various tasks. In contrast, both \textit{Task Overview} and \textit{Task Completed} are brief, consisting of a single acknowledgment turn.
Overall, users spend 68.5\% of their interaction in \textit{Task Grounding}, highlighting the importance of effective task discovery and guidance before entering \textit{Task Execution}.

\subsection{Analyzing User Intents}
\label{sub_sub_user_intents}
\begin{figure}[tb]
    \centering
    \includegraphics[width=0.49\textwidth]{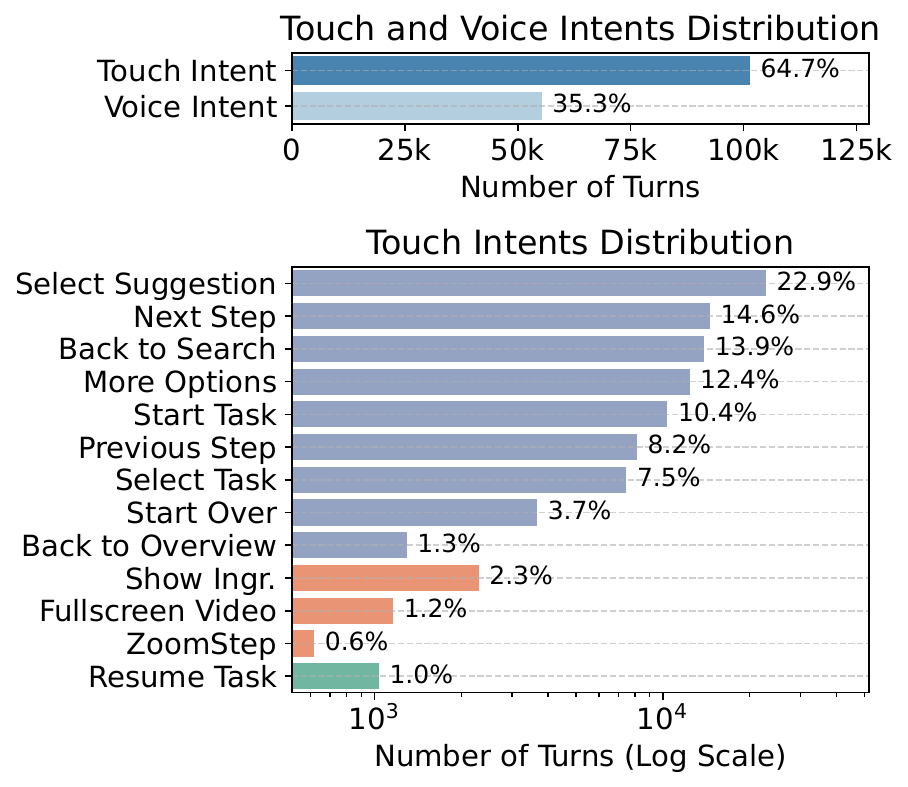}\hfill
    \includegraphics[width=0.49\textwidth]{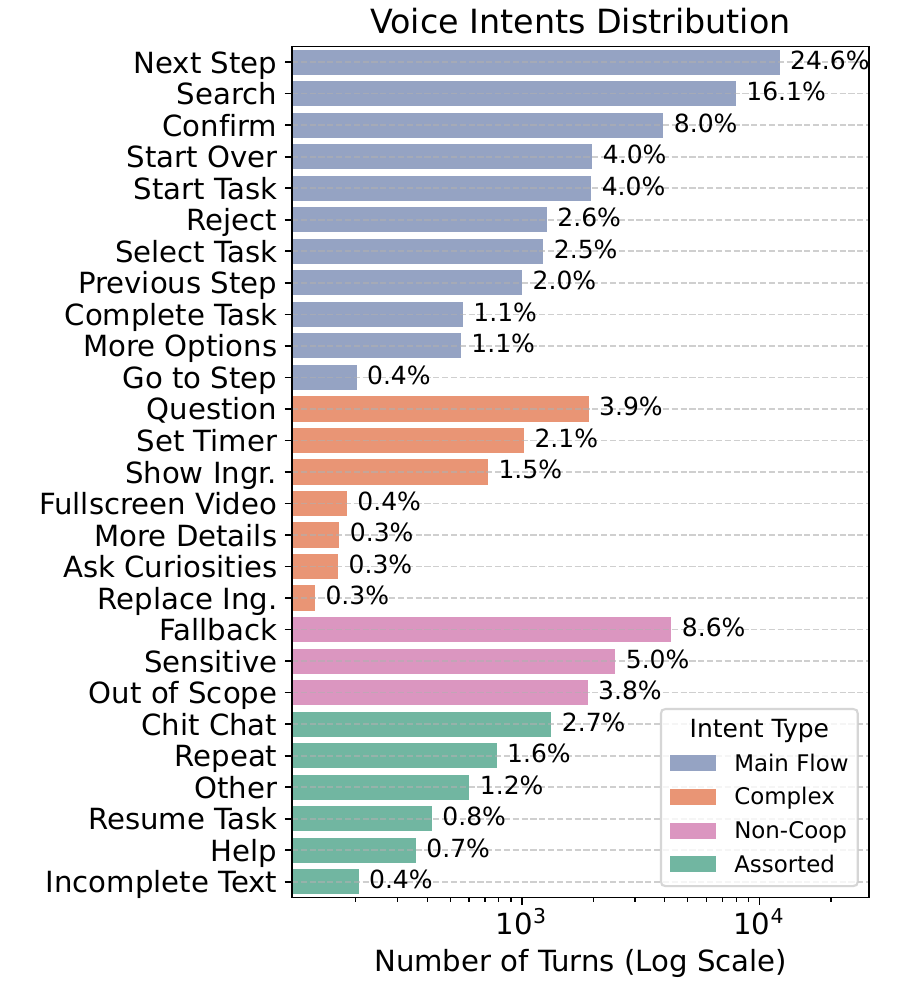}
    \caption{Intent distribution across all dialogues. 
    (top-left) Share of Touch vs. Voice intents by total turns. 
    (bottom-left) Touch intents by type, (right) Voice intents by type. 
    Percentages represent the proportions within each modality.}
    \label{fig_intent_distribution}
\end{figure}

Figure~\ref{fig_intent_distribution} shows the distribution of intents, grouped into \textit{Main Flow} (core navigation), \textit{Complex} (in-depth interactions), \textit{Non-Cooperative} (outside the system domain), and \textit{Assorted} (auxiliary support).

For overall user behavior, we see that 64.7\% of turns use touch controls, likely due to its reliability compared to voice commands~\cite{surpassing_asr_errors}, with most belonging to the main interaction flow. 

For voice-intents, we also see that \textbf{task navigation dominates}, where the most frequent intent is \textit{Next Step}, showing a focus on moving forward in tasks.
Other common voice intents include \textit{Search}, \textit{Confirmation}, and \textit{Start Task}.
Complex intents like \textit{Questions} and \textit{Replace Ingredients} are less frequent, indicating linear task navigation.
We also observe several \textbf{deviations from the main flow}.
\textit{Fallback} intent appears in 8.6\% of turns, indicating unclear input and highlighting opportunities to improve NLU.
\textit{Out-of-Scope} (e.g., smart home control or music playback) and \textit{Sensitive} requests (e.g., sex, politics) are also common, reflecting a presence of non-cooperative~\cite{interaction_non_cooperative_user} behaviors in CTA interactions.

\subsection{Analysis of Conversational User Traits}
\label{sub_sub_user_traits}

\begin{figure*}[tb]
    \centering
    
    % --- TOP ROW: Dialogue-level traits ---
    \includegraphics[width=0.25\textwidth]{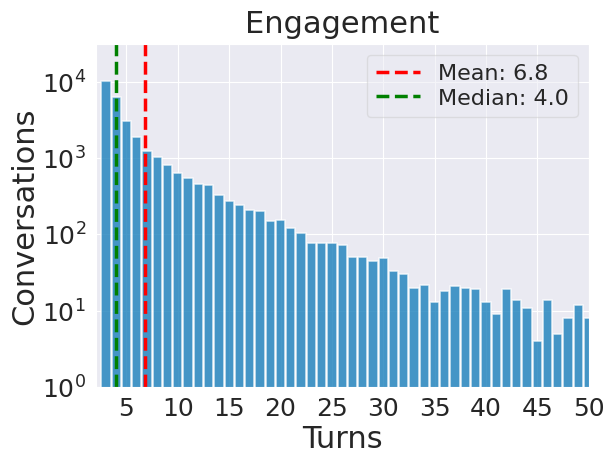}\hfill
    \includegraphics[width=0.25\textwidth]{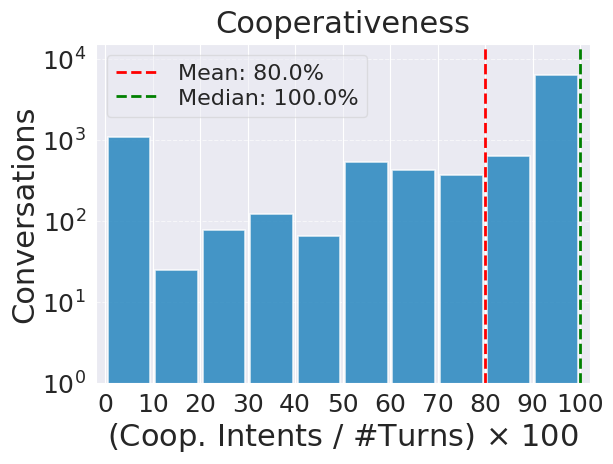}\hfill
    \includegraphics[width=0.25\textwidth]{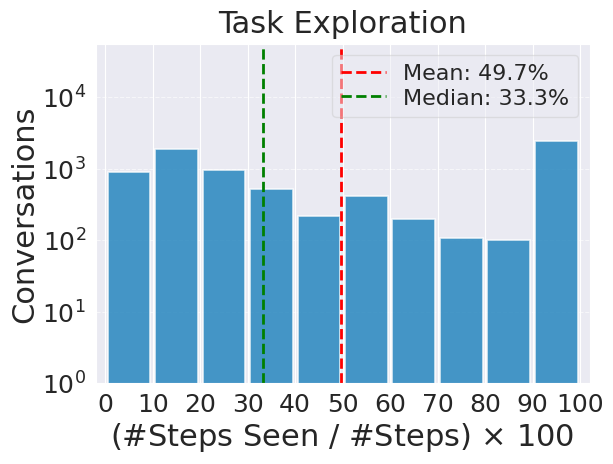}\hfill
    \includegraphics[width=0.25\textwidth]{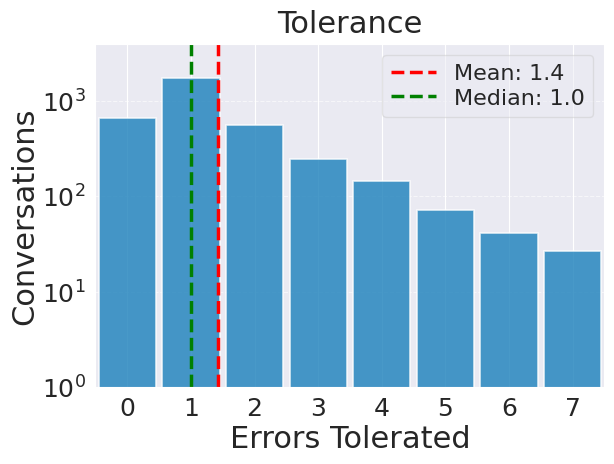}

    % --- BOTTOM ROW: Utterance-level traits ---
    \includegraphics[width=0.25\textwidth]{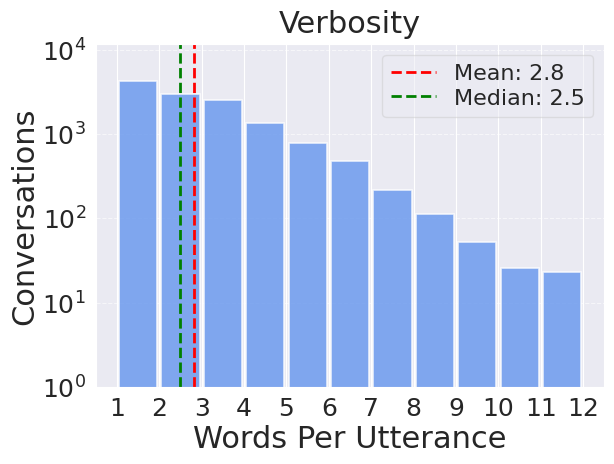}\hfill
    \includegraphics[width=0.25\textwidth]{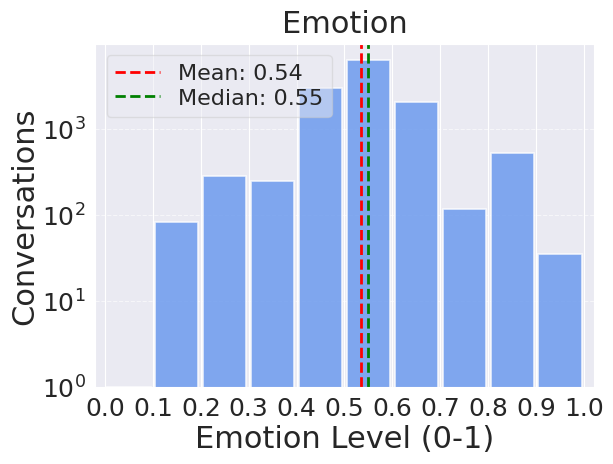}\hfill
    \includegraphics[width=0.25\textwidth]{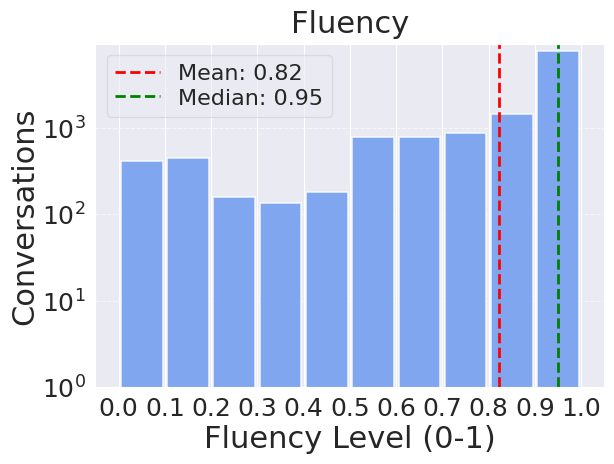}\hfill
    \includegraphics[width=0.25\textwidth]{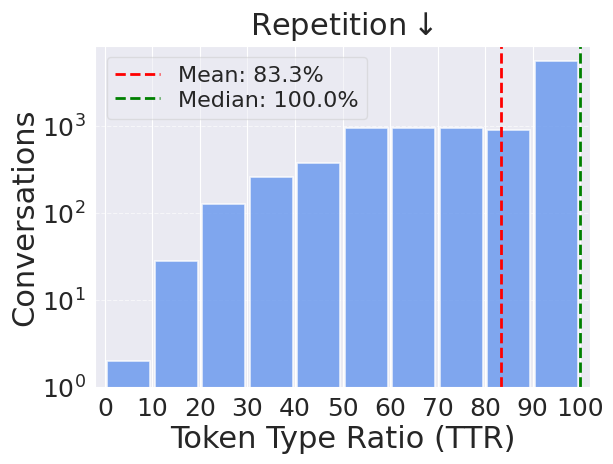}
    
\caption{Dialogue-level (top row) and utterance-level (bottom row) trait statistics. The metrics follow~\cite{rafael_user_simulator} and are shown on the x-axis. 
\href{https://huggingface.co/cardiffnlp/twitter-roberta-base-sentiment-latest}{Emotion level} 
and \href{https://huggingface.co/gchhablani/bert-base-cased-finetuned-cola}{Fluency level} 
are model-based metrics, as in~\cite{rafael_user_simulator}.}
    \label{fig_traits_distribution}
\end{figure*}

To better understand user–CTA dynamics, we analyze conversational behavior through the lens of user traits. Specifically, we adopt the taxonomy of eight user traits defined in~\cite{rafael_user_simulator}, which captures differences in how users interact with conversational systems.
The distribution for each trait is shown in Figure~\ref{fig_traits_distribution}.

Considering \textbf{Dialogue-level} traits which impact the progression of a conversation, we see that 
\textit{Engagement} has an average of 6.8 turns per dialogue, with shorter interactions often indicating user exploration, irrelevant tasks, or early abandonment. \textit{Cooperativeness} is generally high ($\mu=80\%$), though a small peak near zero reflects fully uncooperative users. For \textit{Task Exploration}, users who start a task complete on average half of the steps, with task-length also influencing its completion. \textit{Tolerance} is limited, as users generally accept only one or two system fallbacks before ending the session.

For \textbf{Utterance-level} traits, which influence the style of the
user utterances, we see that 
\textit{Verbosity} is low ($\mu=2.8$ words), reflecting concise task-focused commands~\cite{simple_user_commands_2}. \textit{Emotion} is mostly neutral ($\mu=0.54$), with extremes corresponding to frustration or satisfaction. \textit{Fluency} is generally high ($\mu=0.82$) but drops for longer utterances, often due to ASR errors or hesitation. \textit{Repetition} is low on average, indicating diverse vocabulary usage across turns.

\subsubsection{Real vs. Crowdsourced Behavior.} 
Prior work on CTAs relies on crowdsourced datasets, however, such controlled settings may not reflect real-world usage~\cite{tavares_fake_users}. To quantify this gap, we compare our real user data with the crowdsourced Wizard of Tasks dataset~\cite{wizard_of_tasks} across user traits, revealing substantial differences.
Complex actions occur in only 8.8\% of real interactions versus 47.5\% in crowdsourced data, reflecting limited exploration in natural usage.
Real utterances are shorter (2.9 vs.\ 12.8 words), noisier, and more informal, with uncooperative behavior (e.g., fallbacks, out-of-scope, sensitive requests) appearing in $\approx17\%$ of interactions while being mostly absent from crowdsourced datasets. 
This gap highlights the importance of real-world data for building robust CTAs.

\subsection{User Rating Analysis in CTAs}
\label{sub_results_user_ratings}
\begin{figure}[tb]
\centering
\includegraphics[width=0.84\linewidth]{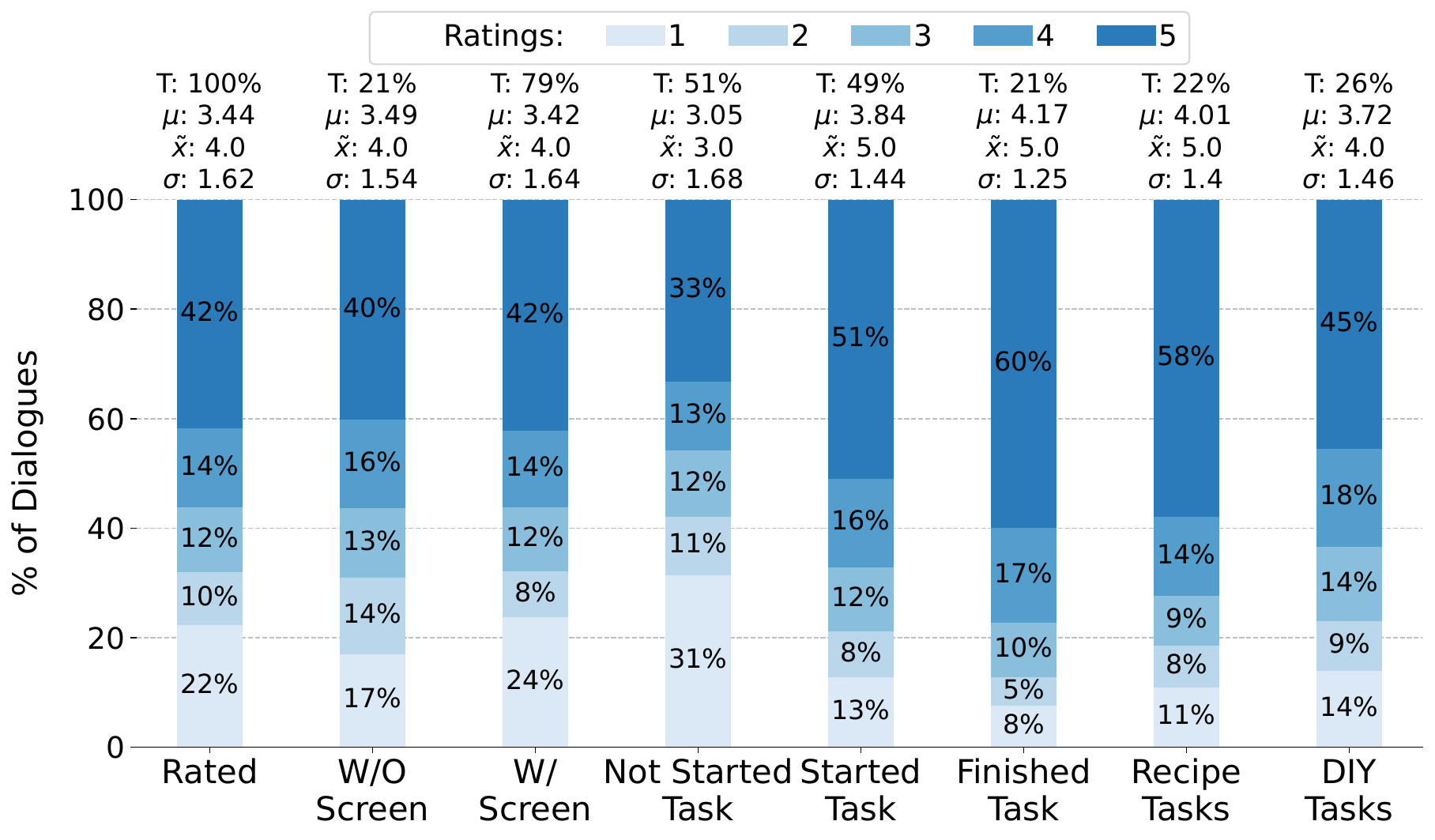}
\caption{Ratings distribution (\%), mean ($\mu$), median ($\tilde{x}$), and std ($\sigma$), across various settings.}
\label{fig_ratings_distribution}
\end{figure}

In Figure~\ref{fig_ratings_distribution}, we analyze the 5k rated conversations, providing insights into system performance and user satisfaction.
The average rating is 3.44, clustering around 1 and 5, indicating users predominantly provide feedback after strong positive or negative experiences, with most rated interactions being positive ($\geq$ 3).

Ratings do not differ substantially between device types (3.49 for non-screen devices vs.\ 3.42 for screen-based devices).
In contrast, engaging with tasks is strongly associated with satisfaction: users who start a task rate the system higher than those who do not (3.84 vs.\ 3.05), with the highest ratings among users who complete tasks (4.17). 
Task domain is also associated with differences in ratings, with recipes scoring higher than DIY tasks (4.01 vs.\ 3.72), likely due to more curated content.

\subsubsection{Behavioral Signals Associated with User Satisfaction.}
\label{sub_sub_rating_correlations}

\begin{figure}[thb]
\centering
\includegraphics[width=0.99\linewidth]{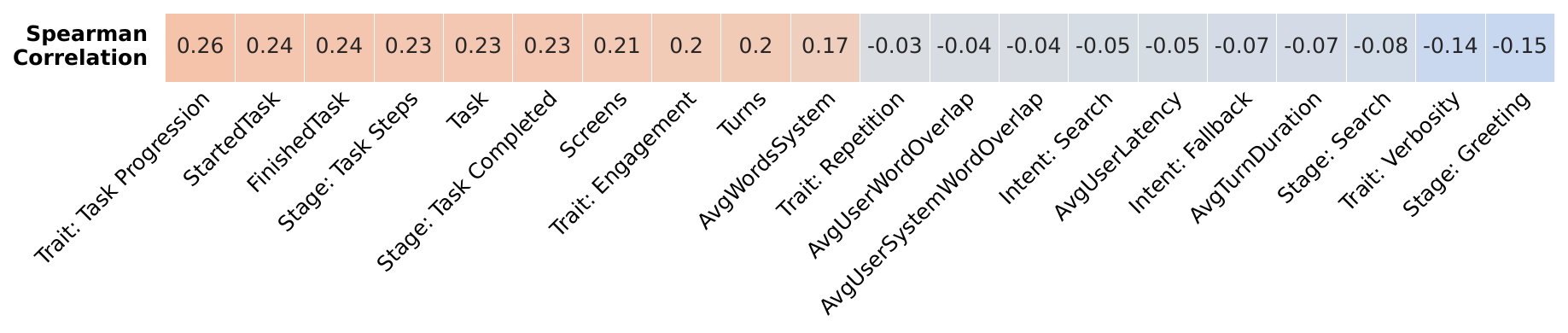}
\caption{Top-20 absolute Spearman correlation coefficients between each feature and user ratings. All results are statistically significant at $p<0.01$.}
\label{fig_ratings_correlation}
\end{figure}

We extend the general features of \textit{Choi et al.}~\cite{offline_online_satisfaction} with CTA-specific signals (intents, task stages, and user traits), resulting in a total of 39 features. 
The Spearman correlation analysis in Figure~\ref{fig_ratings_correlation} shows that no single feature is strongly correlated with user ratings, suggesting that user satisfaction is related to a combination of behavioral and contextual factors~\cite{offline_online_satisfaction, italian_rating_prediction}.

The strongest positive correlations with rating are task-related: \textit{Task Progression} (0.26), \textit{Start} and \textit{Completion} (0.24), and reaching \textit{Task Execution} (0.23). 
These associations indicate that conversations involving task progression tend to receive higher user ratings. Negative correlations are observed for more turns in earlier stages such as \textit{Greeting} (-0.16) and \textit{Search} (-0.08), \textit{User Verbosity} (-0.14), and users taking more time to answer (-0.06), which may reflect greater early-stage friction or user frustration.

\subsubsection{Subjectivity of Ratings.}
\label{sub_sub_ratings_subjectivity}

To evaluate rating subjectivity, we sampled 100 conversations (20 per rating level, $<20\%$ touch input). Following \textit{Cervone et al.}~\cite{italian_rating_prediction}, we asked three annotators to independently rate each transcript on the same 1–5 scale from the user’s perspective, assessing overall satisfaction with the system’s performance.
Ratings were binarized (1–2 negative, 3–5 positive) and evaluated using Spearman correlation ($\rho$), agreement, and Fleiss’ $\kappa$. 

Inter-annotator agreement was high ($\rho=0.70$, 80\%, $\kappa=0.59$), while agreement with user ratings was substantially lower ($\rho=0.24$, 61\%, $\kappa=0.22$). We attribute this to annotators applying stricter evaluation criteria than end users, leading to lower average scores (2.7 vs. 3.4).
These results highlight the variability of user ratings and the difference between experiencing a dialogue and evaluating it post hoc, as also shown by \textit{Choi et al.}~\cite{offline_online_satisfaction}.

\section{Challenges, Opportunities and Design Implications}
\label{sub_sub_common_errors}
After the careful analysis of user-CTA interaction data, we identified multiple opportunities and research challenges associated with lower performing conversations. In addition, we put-forward key design implications for CTA systems.

\subsubsection{Ambiguous ASR and Contexts.}
ASR errors frequently disrupted interactions. Users often attempted standard voice assistant features (e.g., music, smart home) unavailable during the CTA interaction, suggesting the need for multitasking between core assistant features and 3\textsuperscript{rd}-party applications.

\subsubsection{Agent Knowledge Boundaries.}
Task search was the most common problem when content relevant to the user query was not found. Queries requiring external knowledge (e.g., ingredient substitutions or Q\&A) were often associated with incorrect responses, pointing to potential improvements via RAG methods~\cite{rag_survey}.

\subsubsection{User Behavior.}
Most users followed a linear task flow with limited use of complex features such as Q\&A or multimodal interaction. Moreover, the prevalence of uncooperative behavior illustrates challenges, including feature discoverability~\cite{discovering_commands} and proactive support during task execution.

\subsubsection{CTA Design Guidelines.}
\begin{figure}[tb]
\centering
\includegraphics[width=0.92\textwidth]{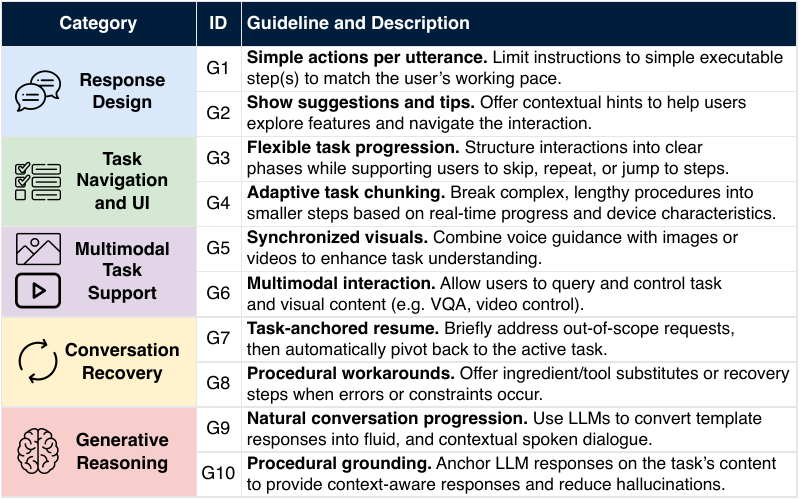}
\caption{Design Guidelines for Conversational Task Assistants.}
\label{tab_guidelines_concise}
\end{figure}

Drawing on insights from our analysis of real-world interactions, we synthesize our findings into a set of design guidelines for more behavior-driven and robust CTAs, summarized in Figure~\ref{tab_guidelines_concise}.

\section{Conclusions}
\label{sec_conclusions}

We presented a large-scale study of a CTA, TWIZ-v2, analyzing real-world interactions to characterize user behavior, conversational traits, and user satisfaction.
Our findings highlight the importance of real-world data, revealing various interaction patterns, including preferences for touch-based interaction, predominantly linear task progression, and frequent out-of-domain behavior. We further show that user satisfaction is associated with engagement in task execution.

In summary, these results highlight key opportunities for improving real-world CTAs, including enhancing robustness, task discovery, system knowledge, and error recovery, and provide empirically grounded and actionable guidelines for designing more adaptive and context-aware CTAs.

\section*{Limitations}
\label{sec_limitations}

The analyzed conversations originate from U.S. Alexa users interacting in English. Although the U.S. user base is large and diverse, the observed interaction patterns may differ across other regions, languages, cultural backgrounds, and platforms. Furthermore, while the system supports a wide range of cooking and DIY tasks, the findings may not directly generalize to all task domains. Additionally, the reported associations with ratings are correlational, and may reflect unmeasured factors such as user motivation or task relevance. Investigating these settings are avenues for future work.

\section*{Ethical Considerations}

All data used in this study were collected from users who voluntarily interacted with the system and consented to the use of their interactions for research purposes. All conversations are fully anonymized, with no personally identifiable information retained, and all results are reported only in aggregate form.

\section*{Acknowledgements}
This work has been partially funded by the Amazon Science - TaskBot Prize Challenge 2021 and 2022, by the NOVA LINCS project Ref. UIDP/04516/2020 and by FCT Ref. UI/BD/151261/2021.

\bibliographystyle{splncs04}
\bibliography{custom}

\end{document}